\documentclass{PoS}

\title{
{\protect
\vspace{-2cm}
\hspace*{11cm}{
\begin{minipage}{0.34\hsize}
\normalsize
UTCCS-P-35\\
HUPD-0706 \\
KANAZAWA-07-17\\
\end{minipage}} \\
\vspace{1cm}
}
Light hadron spectrum with 2+1 flavor dynamical $O(a)$-improved Wilson quarks}

\ShortTitle{Light hadron spectrum with 2+1 flavor dynamical $O(a)$-improved Wilson quarks}

\author{PACS-CS Collaboration: 
\speaker{N. Ukita}${}^{a}$\thanks{E-mail: ukita@ccs.tsukuba.ac.jp},
 S.~Aoki${}^{b,c}$,
 N. Ishii${}^{a}$,
 K.-I.~Ishikawa${}^{d}$,
 N.~Ishizuka${}^{a,b}$,
 T. Izubuchi${}^{c,e}$,
 D. Kadoh${}^{a}$,
 K.~Kanaya${}^{b}$,
Y. Kuramashi${}^{a,b}$,
Y. Namekawa${}^{a}$,
 M.~Okawa${}^{d}$,
 K. Sasaki.${}^{a}$,
 Y.~Taniguchi${}^{a,b}$,
 A.~Ukawe${}^{a,b}$,
 T.~Yoshi\'e${}^{a,b}$
 \\
 \llap{${}^a$}Center for Computational Sciences, University of Tsukuba, Tsukuba, Ibaraki 305-8577, Japan\\
 \llap{${}^b$}Graduate School of Pure and Applied Sciences, University of Tsukuba, Tsukuba, Ibaraki 305-8571, Japan\\
 \llap{${}^c$}Riken BNL Research Center, Brook-haven National Laboratory, Upton, New York 11973, USA\\
 \llap{${}^d$}Department of Physics, Hiroshima University, Higashi-Hiroshima, Hiroshima 739-8526, Japan\\
 \llap{${}^e$}Institute for Theoretical Physics, Kanazawa University, Kanazawa, Ishikawa 920-1192, Japan}
\abstract{We present preliminary results for the light  harden spectrum in $N_f=2+1$ lattice QCD obtained with the nonperturbatively $O(a)$-improved Wilson quark action and the
Iwasaki gauge action. 
Simulations are carried out at $\beta=1.90$ on a $32^3 \times 64$
lattice using the PACS-CS computer. 
We employ L\"uscher's domain-decomposed HMC
algorithm to reduce the up-down quark masses toward the physical
value. The pseudoscalar meson masses range from 730 MeV down to 210 MeV. 
We compare the light harden spectrum extrapolated to the physical point
with the experimental values.}

\FullConference{The XXV International Symposium on Lattice Field Theory
		 July 30-4 August 2007\\
		 Regensburg, Germany}

\begin{document}

\section{Introduction}

It is an essential step for lattice QCD calculation to
reproduce the harden spectrum with various systematic errors under full control
and hence show that QCD is the fundamental theory of the strong interaction.
Recent developments in simulation algorithms and computational
facilities now allow us to remove the most troublesome systematic
errors: the quenching effect and chiral extrapolation. 
The main task of the previous CP-PACS/JLQCD project \cite{cppacs/jlqcd1,
cppacs/jlqcd2} was to remove the former by
performing $N_f=2+1$ lattice QCD simulations 
with the nonperturbatively $O(a)$-improved Wilson quark action \cite{oa}
and the Iwasaki gauge action \cite{iwasaki} 
on a $(2\rm fm)^3$ lattice at three lattice spacings.
The lightest up-down quark mass, however, was still 64MeV corresponding to
$m_{\pi}/m_{\rho}\approx0.6$ so that there remained an ambiguity from a 
long chiral extrapolation.

In the PACS-CS(Parallel Array Computer System for Computational
Science) project \cite{ukawa1, ukawa2, kuramashi, kuramashi2}, 
we aim to perform $N_f=2+1$ lattice QCD simulations at the physical 
point using the PACS-CS computer with a total peak speed of 14.3 TFLOPS 
developed and installed at University of Tsukuba on 1 July 2006. 
The quark and gauge actions are the same as 
in the previous CP-PACS/JLQCD project.  
To reduce the degenerate up and down quark masses 
we employ the domain-decomposed HMC (DDHMC) algorithm 
with the replay trick proposed by L\"uscher \cite{luscher, kennedy},
whose effectiveness for small quark mass region 
has already been shown in the $N_f=2$ case \cite{luscher, del}
incorporating the multiple time scale integration
scheme \cite{sexton}.
The strange quark is included 
by the UV-filtered Polynomial Hybrid Monte Carlo (UV-PHMC) 
algorithm \cite{ishikawa}. 

In this report we present simulation details and some preliminary results  
for the hadron spectrum. 
The analysis on the chiral behavior of the pseudoscalar meson masses
and the decay constants using chiral perturbation theory
is given in a separate report~\cite{kadoh}.

\section{Simulation details}
\begin{table}[b!] 
\setlength{\tabcolsep}{10pt}
\renewcommand{\arraystretch}{1.2}
\centering
%\begin{tabular}{ccccccccc}% \hline
% $\beta=1.90$, $L^3\times T=32^3\times64$, $c_{\rm SW}=1.715$\\ %\hline
%\end{tabular}
\begin{tabular}{cccccccc} \hline
$\kappa_{\rm ud}$ & $\kappa_{\rm s}$ &$\tau$& $(N_0,N_1,N_2)$ & $N_{\rm poly}$ & MD time & $\tau_{\rm int}[P]$ \\ \hline \hline
0.13700 & 0.13640 &0.5& (4,4,10) &180& 2000 & 38.2(17.3)\\
0.13727 & 0.13640 &0.5& (4,4,14) &180& 2000 & 20.9(10.2)\\
0.13754 & 0.13640&0.5&(4,4,20) &180& 2500 & 19.2(8.6)   \\
0.13770 & 0.13640 &0.25& (4,4,16) &180& 2000& 38.4(25.2)\\
0.13781 & 0.13640 &0.25& (4,4,48) &180& 350 &9.1(6.1)   \\ \hline
0.13754 & 0.13660 &0.5& (4,4,28)&220& 900 & 10.3(2.9)   \\ \hline
\end{tabular}
\caption{Summary of simulation parameters. MD time is the number of
trajectories multiplied by the trajectory length $\tau$.
$\tau_{\rm int}[P]$ denotes the integrated autocorrelation time for 
the plaquette.}
\label{tab:param}
\end{table} 

We employ the $O(a)$-improved Wilson quark action with a nonperturbative
improvement coefficient $c_{\rm SW}=1.715$\cite{csw} and 
the Iwasaki gauge action at $\beta=1.90$ on a $32^3\times64$ lattice. 
%The lattice spacing is about $0.09$fm.  
Table~\ref{tab:param} summarizes our simulation parameters.
Simulations are carried out for six combinations of 
the hopping parameters $(\kappa_{\rm ud}, \kappa_{\rm s})$. 
The heaviest case of $(\kappa_{\rm ud}, \kappa_{\rm s})=(0.13700,
0.13640)$ is chosen for a direct comparison of the PACS-CS results
with the previous CP-PACS/JLQCD ones which employed
this parameter as the smallest quark mass.    
Among two choices of $\kappa_{\rm s}$, $\kappa_{\rm s}=0.13640$ is  
the physical point $\kappa_{\rm s}=0.136412(50)$ as determined 
in the previous CP-PACS/JLQCD work\cite{cppacs/jlqcd1, cppacs/jlqcd2}, 
while $\kappa_{\rm s}=0.13660$ is to investigate the strange 
quark mass dependence. 

The DDHMC algorithm is implemented by domain-decomposing the full
lattice with a $8^4$ block size as a preconditioner for HMC. 
The domain-decomposition factorizes the up-down quark determinant to
 the UV and the IR parts distinctively.
On the other hand, the UV-PHMC algorithm for the strange quark 
is not domain-decomposed. 
We can incorporate the multiple time scale 
integration scheme in these algorithms
to reduce the simulation cost efficiently.
The relative magnitudes of the force terms are found to be 
\begin{eqnarray}   
||F_{\rm g}||:||F_{\rm UV}||:||F_{\rm IR}|| \approx 16:4:1,
\end{eqnarray}
where $F_{\rm g}$ denotes the gauge part and $F_{\rm UV, IR}$ for the UV
and the IR parts of the up-down quarks.
We write the associated step sizes for the above forces as 
$\delta\tau_{\rm g}=\tau/N_0 N_1 N_2,\ \  \delta\tau_{\rm UV}=\tau/N_1
N_2,\ \  \delta\tau_{\rm IR}=\tau/N_2$ with $\tau$ the trajectory length, 
where the integers $N_0, N_1,  N_2$ are chosen 
such that 
\begin{eqnarray}
 \delta\tau_{\rm g} ||F_{\rm g}|| \approx \delta\tau_{\rm UV} ||F_{\rm UV}|| \approx \delta\tau_{\rm IR} ||F_{\rm IR}||.
 \end{eqnarray} 
We take $N_0=N_1=4$.
Since we have found $||F_{\rm s}||\approx ||F_{\rm IR}||$ 
for the strange quark part, we choose   
$\delta\tau_{\rm s}=\delta\tau_{\rm IR}$.
%We choose the trajectory length of $\tau=0.25/\sqrt{2}$ 
%for $\kappa_{\rm ud}=0.13770, 0.13781$,
%and $\tau=0.5/\sqrt{2}$ for other $\kappa_{\rm ud}$. 
The value of $N_2$ and the polynomial order 
for UV-PHMC $N_{\rm poly}$ are adjusted 
taking account of acceptance rate and simulation stability.

The inversion of the Wilson-Dirac operator $D$ is carried out by the GCR
solver with the stopping condition $|Dx-b|/|b|<10^{-9}$ for the force
calculation and $10^{-14}$ for the Hamiltonian, which guarantees 
the reversibility of the molecular dynamics trajectories to high
precision: $|\Delta U|<10^{-12}$ for the link variables and 
 $|\Delta H|<10^{-8}$ for the Hamiltonian. 
 
\section{Plaquette history and autocorrelation time}
\begin{figure}[b!]
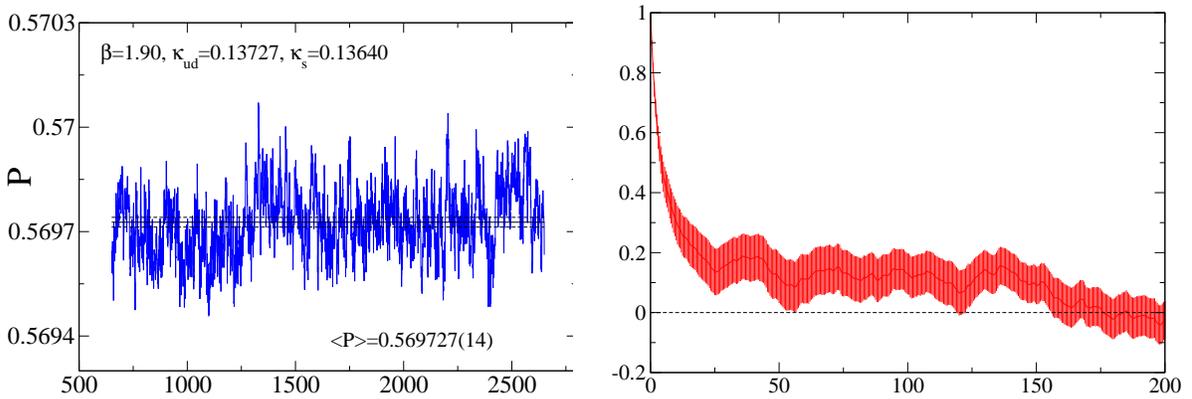

\vspace{3mm}
\begin{center}
\begin{tabular}{cc}
\includegraphics[width=76mm,angle=0]{./fig/PLQ_his_b190kud013727ks013640b.eps}  &
\includegraphics[width=76mm,angle=0]{./fig/PLQ_autocorr.eps} 
\end{tabular}
\end{center}
\vspace{-.5cm}
\caption{Plaquette history (left) and normalized autocorrelation 
function (right) for $(\kappa_{\rm ud}, \kappa_{\rm s})=(0.13727, 0.13640)$. }
\label{fig:PLQ}
\end{figure}

In Fig.~\ref{fig:PLQ} we show the plaquette history and the normalized 
autocorrelation function at
$(\kappa_{\rm ud}, \kappa_{\rm s})=(0.13727, 0.13640)$ as a
representative case.
The integrated autocorrelation time 
is estimated as $\tau_{\rm int}[P]=20.9(10.2)$
following the definition in Ref.~\cite{luscher}. 
For other hopping parameters 
we have found similar behaviors for the normalized autocorrelation 
function. Although we hardly observe the quark mass dependence for 
$\tau_{\rm int}[P]$ listed in Table~\ref{tab:param},
the statistics may not be
sufficiently large to derive a definite conclusion.

\section{Hadron spectrum}

We measure hadron correlators at every 10 trajectories
at the unitary points where 
the valence quark masses are equal to the sea quark masses.
Light hadron masses are extracted from single exponential 
$\chi^2$ fits to the correlators
with an exponentially smeared source and a local sink. 
Statistical errors are estimated by the jackknife method, whose
bin size is chosen to be 50 molecular dynamics time  
based on the bin size dependence of the errors.

%\section{Comparison with the CP-PACS/JLQCD results}
We first compare the PACS-CS results on $32^3\times 64$ 
with the previous CP-PACS/JLQCD results 
on $20^3\times 40 $\cite{cppacs/jlqcd1, cppacs/jlqcd2} at $\beta=1.90$.
%at $(\kappa_{\rm ud}, \kappa_{\rm s})=(0.13700, 0.13640)$.
Figure~\ref{fig:comp} shows the up-down quark mass dependence of
$(am_{\pi})^2$ and $am_{\rho}$ with $\kappa_{\rm s}$ fixed at 0.13640.
As for the pion mass we observe that the PACS-CS and 
the CP-PACS/JLQCD results are smoothly connected 
as a function of $1/\kappa_{\rm ud}$.
Table~\ref{tab:comp} shows that the results for 
the pion mass at $\kappa_{\rm ud}=0.13700$
are consistent within the error bars.
On the other hand, we find 1$-$2\% deviation 
for the $\rho$ meson and nucleon masses, which
could be finite size effects.

\begin{figure}[t!]
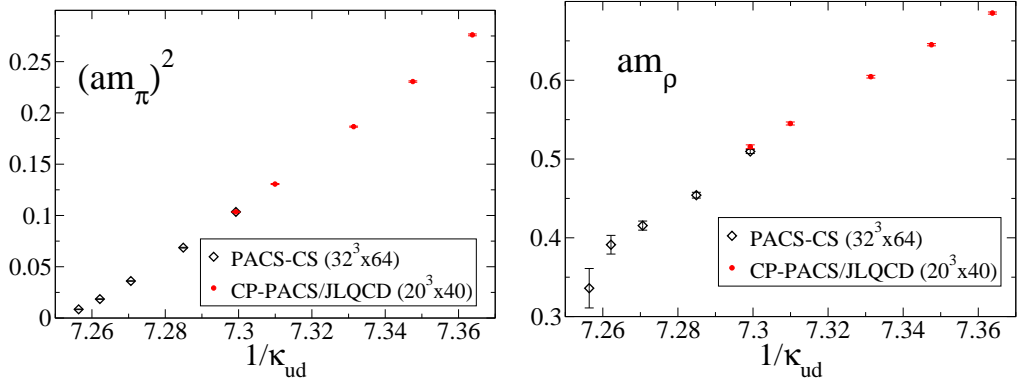

\vspace{3mm}
\begin{center}
\begin{tabular}{cc}
\includegraphics[width=65mm,angle=0]{./fig/kud.PS_LL2.eps} &
\includegraphics[width=65mm,angle=0]{./fig/kud.V_LL.eps}
\end{tabular}
\end{center}
\vspace{-.5cm}
\caption{$(am_{\pi})^2$ (left) and $am_{\rho}$ (right) 
as a function of $1/\kappa_{\rm ud}$. 
Black and red symbols denote the PACS-CS and the CP-PACS/JLQCD results, 
respectively.}
\label{fig:comp}
\end{figure}

\begin{table}[t!]
\centering
\begin{tabular}{cccccccc}  \hline
                                 & lattice size           &  $am_{\pi}$ & $am_{\rho}$ & $am_{\rm N}$\\ \hline
PACS-CS               &$32^3\times 64$& 0.3220(6) & 0.506(2) & 0.726(3) \\ 
CP-PACS/JLQCD&$20^3\times 40$ &0.3218(8)& 0.516(3) & 0.733(4)\\ \hline
\end{tabular}
\caption{PACS-CS and CP-PACS/JLQCD results for 
$\pi$, $\rho$ and nucleon masses at 
$(\kappa_{\rm ud}, \kappa_{\rm s})=(0.13700, 0.13640)$.}
\label{tab:comp}
\end{table}

\begin{figure}[t!]
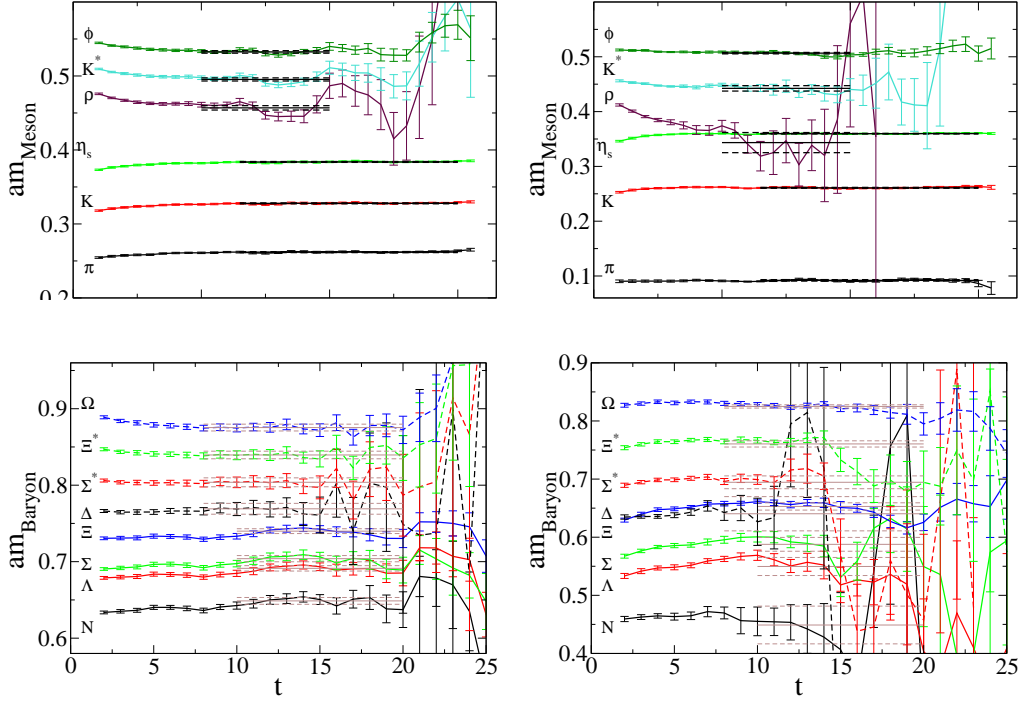

\vspace{3mm}
\begin{center}
\begin{tabular}{cc}
\includegraphics[width=65mm,angle=0]{./fig/msn_13727.eps} &
\includegraphics[width=65mm,angle=0]{./fig/msn_13781.eps} \\
\includegraphics[width=65mm,angle=0]{./fig/brn_13727.eps} &
\includegraphics[width=65mm,angle=0]{./fig/brn_13781.eps}
\end{tabular}
\end{center}
\vspace{-.5cm}
\caption{Effective masses for the mesons (top) and the baryons
 (bottom) at $\kappa_{\rm ud}=0.13727$ (left) and 0.13781 (right).}
\label{fig:Meff}
\end{figure}

\begin{figure}[t!]
\vspace{3mm}
\begin{center}
\begin{tabular}{cc}
\includegraphics[width=75mm,angle=0]{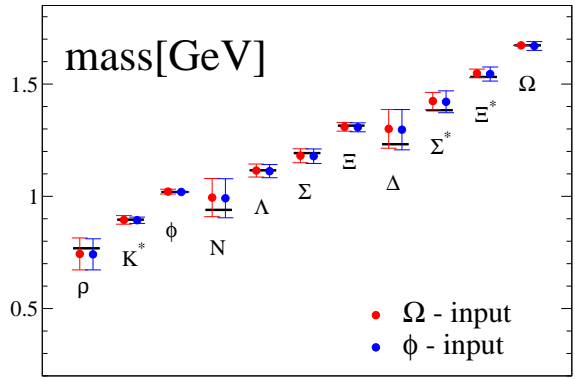}
\end{tabular}
\end{center}
\vspace{-.5cm}
\caption{Light hadron spectrum extrapolated at the physical point 
with  $\Omega$-input (red) and $\phi$-input (blue).}
\label{fig:comp}
\end{figure}

Figure~\ref{fig:Meff} shows the hadron effective masses  
at $\kappa_{\rm ud}=0.13727$ and 
$0.13781$. We observe clear plateau for the mesons except for the $\rho$
meson and also good signal for the baryons thanks to a large volume.
Especially, the $\Omega$ baryon has a stable signal 
and a weak up-down quark mass dependence for our simulation parameters.   
Taking advantage of this virtue we employ the 
$\Omega$ baryon as input to determine the lattice cutoff.
Combined with the additional inputs of $m_\pi$ and $m_K$ to determine
the physical up-down and strange quark masses,
we obtain $a^{-1}=2.256(80)$GeV. 
With the use of this cutoff we find that the lightest
pseudoscalar meson mass we have reached is about 210MeV.
The light hadron spectrum extrapolated to the physical point is shown
in Fig.~\ref{fig:comp}, and compared with the experimental values.
They show a good agreement within the
error bars, though our results contain
possible $O((a\Lambda_{\rm QCD})^2)$ cutoff errors.
In Fig.~\ref{fig:comp} we also plot the results for the $\phi$-input
case, which are consistent with the $\Omega$-input case.  
A more detailed description about the determination of $a^{-1}$ and the
chiral extrapolation of the hadron masses is given in Ref.~\cite{kadoh}.

We calculate the bare quark masses using the axial vector
Ward-Takahashi identity (AWI) defined by
$am^{\rm AWI}=\lim_{t\rightarrow \infty}
{\langle \nabla_4A_4^{\rm imp}(t)P(0) \rangle}/(2\langle P(t)P(0)\rangle)$,
where $A_4^{\rm imp}$ is the nonperturbatively 
$O(a)$-improved axial vector current \cite{A4I}.
Employing the perturbative renormalization factors $Z_{A,P}$ 
evaluated up to one-loop level \cite{Z1, Z2}, we obtain
\begin{eqnarray}
m_{\rm ud}^{\overline{\rm MS}}(\mu=2{\rm GeV})=2.3(11){\rm MeV},&&
m_{\rm s}^{\overline{\rm MS}}(\mu=2{\rm GeV})=69.1(25){\rm MeV}.
\end{eqnarray}
The physical up-down quark mass is about half of
our lightest one $m_{\rm ud}^{\overline{\rm MS}}(\mu=2{\rm GeV})=5.61(40)$ 
at $(\kappa_{\rm ud}, \kappa_{\rm s})=(0.13781, 0.13640)$.
We also measure the pseudoscalar meson decay constants, for which 
we find
\begin{eqnarray}
f_\pi=144(6){\rm MeV}, \ \ & f_K=175(6){\rm MeV}, \ \ & f_K/f_{\pi}=1.219(22)
\end{eqnarray}
at the physical point with the perturbative $Z_A$.
They are about 10\% larger than the experimental values, though their  ratio, which
is free from the ambiguities of the renormalization factors, is consistent with
the experimental value. 
For both the quark masses and the pseudoscalar meson decay constants,
our concern is the use of the perturbative renormalization factors
which might cause sizable systematic errors. 
We are now calculating the nonperturbative $Z_{A,P}$ with the
Schr{\"o}dinger functional scheme.

\section{$\rho$-$\pi\pi$ mixing}
\begin{figure}[t!]
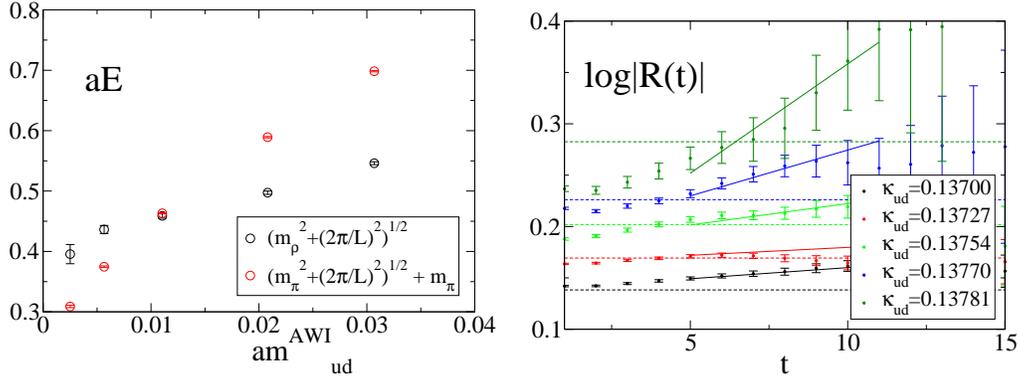

\vspace{3mm}
\begin{center}
\begin{tabular}{cc}
\includegraphics[width=65mm,angle=0]{./fig/E-level_p1.eps}  &
\includegraphics[width=65mm,angle=0]{./fig/log.msn_p1_k1_s11_rho_cpca_sy.eps}
\end{tabular}
\end{center}
\vspace{-.5cm}
\caption{Energy levels of the $\rho$ meson and the two pion states
with the momentum $2\pi/L$ as a function of the
up-down quark mass with $\kappa_s=0.13640$ (left) 
and time dependence of the $R$ function (right).}
\label{fig:mixing}
\end{figure}

Since our simulations are carried out at sufficiently 
small up-down quark masses, 
%to satisfy the condition $m_\rho < 2m_\pi$,
it would be interesting to investigate the $\rho$-$\pi\pi$ mixing effects \cite{bernard, McN}.
We find that, while the rest mass $m_\rho$ is always smaller than 
the two-pion energy $2\sqrt{m_\pi^2+(2\pi/L)^2}$ for all the hopping parameters,
the energy of a moving $\rho$ with a unit of momentum, {\it i.e.,} 
$\sqrt{m_\rho^2+(2\pi/L)^2}$, becomes larger than
$\sqrt{m_\pi^2+(2\pi/L)^2}+m_\pi$ toward the lighter up-down quark
masses. This situation is illustrated in Fig.~\ref{fig:mixing}

Let us consider two types of the $\rho$ meson propagator with the
momentum $2\pi/L$: $\rho_\|(2\pi/L)$ with polarization parallel
to the spatial momentum and $\rho_\bot(2\pi/L)$ with polarization 
perpendicular to the spatial momentum.
Phenomenologically the $\rho$-$\pi\pi$ coupling is described by
$g_{\rho\pi\pi}\epsilon_{abc}\rho_\mu^a\pi^a\partial_\mu\pi^c$,
which favors $\rho_\|(2\pi/L)\rightarrow \pi(2\pi/L)\pi(0)$
to $\rho_\bot(2\pi/L)\rightarrow \pi(2\pi/L)\pi(0)$.
We expect that the $\rho_\|(2\pi/L)$ propagator is more
strongly affected by the mixing effects than
the $\rho_\bot(2\pi/L)$ correlator.
Since the mixing effects push up the upper energy level further and
push down the lower energy level,
they could be detected by measuring the $R$ function defined by
\begin{eqnarray}
R(t)=\frac{\langle\rho_\|({\vec p},t)\rho_\|^\dagger({\vec p},0)\rangle}
{\langle\rho_\bot({\vec p},t)\rho_\bot^\dagger({\vec p},0)\rangle}
\stackrel{{\rm large\;\;}t}{\longrightarrow}
Z{\rm e}^{-(E_{\rho_\|}-E_{\rho_\bot})t}.
\label{eq:ratio}
\end{eqnarray} 
In Fig.~\ref{fig:mixing} we plot $\log|R(t)|$ as a function of $t$.
The data show clear positive slopes which indicate 
$E_{\rho_\|}<E_{\rho_\bot}$. We also observe that the magnitude of the
energy difference is rather small for $\kappa_{\rm ud}\le 0.13754$, while
it grows rapidly as the up-down quark mass is reduced for  
$\kappa_{\rm ud} > 0.13754$.
This feature may suggest that the 
$\langle\rho_\|({\vec p},t)\rho_\|^\dagger({\vec p},0)\rangle$ 
correlator is getting dominated by the $\pi\pi$ state toward
the smaller up-down quark masses.
In order to obtain a definite conclusion, we 
need more detailed investigations with increased statistics.

%\section{Summary}
%We have presented a status report on the PACS-CS project which aim at 
%a 2+1 flavor lattice QCD simulation toward the physical point
%on a (2.8)fm$^3$ box using the $O(a)$-improved Wilson quarks. 
%With the aid of the DDHMC algorithm for the up-down quarks we have 
%reached $m_{\pi}=210$MeV which roughly corresponds to 
%$m_{\rm ud}^{\overline{\rm MS}}(\mu=2{\rm GeV})=5.6$MeV. 
%Thanks to the enlarged volume 
%compared to the previous CP-PACS/JLQCD work,
%we obtain good signals not only for the meson masses 
%but also for the baryon masses. 
%Our results for the hadron spectrum at the physical point
%show a good agreement with the experimental values.
%  
%At present we have just started the simulation at the physical point.
%We also calculate the nonperturbative renormalization factors
%for the quark masses and the pseudoscalar meson decay constants.
%Once these calculations are accomplished,
%the next step is to investigate the finite size effects 
%at the physical point and reduce the discretization errors employing
%finer lattice spacings. 

\vspace{5mm}
\noindent
{\bf \large Acknowledgment}

This work is supported in part by Grants-in-Aid for Scientific Research
from the Ministry of Education, Culture, Sports, Science and Technology
(Nos. 13135204, 15540251, 17340066, 17540259, 18104005, 18540250, 18740130, 18740139).

\end{document}